\documentclass{article}

\def\ignore#1{}

\def\psfancypar#1#2{\begingroup\def\par{\endgraf\endgroup\lineskiplimit=0pt}
               \setbox2=\hbox{\large\sc #2}
               \newdimen\tmpht \tmpht \ht2 \advance\tmpht by \baselineskip
               \font\hhuge=Times-Bold at \tmpht
               \setbox1=\hbox{{\hhuge #1}}
               \count7=\tmpht \count8=\ht1
               \divide\count8 by 1000 \divide\count7 by \count8 
               \tmpht=.001\tmpht\multiply\tmpht by \count7 
               \font\hhuge=Times-Bold at \tmpht
               \setbox1=\hbox{{\hhuge #1}}
               \noindent
                \hangindent1.05\wd1
               \hangafter=-2 {\hskip-\hangindent
               \lower1\ht1\hbox{\raise1.0\ht2\copy1}%
                \kern-0\wd1}\copy2\lineskiplimit=-1000pt}

\def\boxit#1{\vbox{\hrule\hbox{\vrule\kern3pt
        \vbox{\kern3pt#1\kern3pt}\kern3pt\vrule}\hrule}}

\def\reals{ { {\rm  I \kern-0.15em R }  } }
\def\complex{ {\,{{\rm C} \kern-0.50em \raise0.20ex {  |}}\, }}

\def\Rbf{{\bf R}}

\def\Nc{{\cal N}}

\def\be{\vskip .3cm \begin{equation}}
\def\ee{\end{equation} \vskip .4cm \noindent}
\def\defeq{{\stackrel{\Delta}{=}}}
\def\Rxx{\Rbf_{\ssstyle X\kern-.1em X}}

\let\ssstyle=\scriptscriptstyle

\def\Kout{\setbox1=\hbox{\Huge\bf K}\hbox to
1.05\wd1{\hspace{.05\wd1}
}}
\def\Sout{\setbox1=\hbox{\Huge\bf S}\hbox to 1.05\wd1{\hspace{.05\wd1}
}}

\input setup
\usepackage{spconf,amsmath,epsfig,epsf,psfrag,amssymb,amsfonts,latexsym, amsmath}
\usepackage{verbatim}
\usepackage[mathscr]{eucal}

\newcommand{\Xmsc}{\mathscr{X}}

\newcommand{\beq}{\begin{equation}}
\newcommand{\eeq}{\end{equation}}

\newcommand{\Var}{\mbox{${\mbox{Var}}$}}

\def\nn{\nonumber}
\def\defeq{\stackrel{\Delta}{=}}
\def\Ebb{{\mathbb E}}

\title{A Large Deviations Approach to Sensor Scheduling for\\[0.2em] Detection of Correlated Random Fields}

\name{Youngchul Sung, Lang Tong, and H. Vincent Poor\thanks{\scriptsize Y. Sung and L.Tong  are with the School of Electrical and Computer Engineering, Cornell University, Ithaca, NY 14853.
 Email:\{ys87,ltong\}@ece.cornell.edu.  H. V. Poor is with Dept. of Electrical Engineering, Princeton University,    Princeton, NJ 08544. Email:poor@princeton.edu.}
\thanks{This work was supported in part by
the Multidisciplinary University Research Initiative (MURI)  under
the Office of Naval Research Contract N00014-00-1-0564.  Prepared through collaborative participation in the Communications and
Networks Consortium sponsored by the U.~S. Army Research Laboratory under
the Collaborative Technology Alliance Program, Cooperative Agreement
DAAD19-01-2-0011. The work of H.V. Poor was supported in part by the Office
of Naval Research Grant N00014-03-1-0102.}}
\address{}
\vspace{2em}

\begin{document}
\maketitle

\ninept
{\footnotesize
\begin{abstract}
The problem of scheduling sensor transmissions for the detection of correlated random fields using spatially deployed sensors is considered.  Using the large deviations principle, a closed-form expression for the error exponent of the miss probability is given as a function of the sensor spacing and signal-to-noise ratio (SNR).  It is shown that the error exponent has a distinct characteristic:  at high SNR, the error exponent is monotonically increasing with respect to sensor spacing, while at low SNR there is an optimal spacing for scheduled sensors.
\end{abstract}
}

\vspace{-0.5em}
\section{Introduction}
\label{sec:intro}
One of the most critical design constraints for large scale sensor networks
is energy efficiency.   In the context of detecting
spatially correlated signals,  this means that, given
a desired level of detector performance,
one should minimize the amount of data required, and
collect data from judiciously chosen areas.

\begin{figure}[htbp]
\centerline{
    \begin{psfrags}
    \psfrag{0}[c]{ $0$}
    \psfrag{L}[c]{ $x_n$}
    \psfrag{a}[r]{ $s(x)$}
    \psfrag{d}[c]{ $\Delta_{i}$}
    \psfrag{xi}[c]{ $x_i$}
    \psfrag{xi1}[c]{ $x_{i+1}$}
    \epsfxsize=2.7in
    \epsfysize=0.9in
    \epsfbox{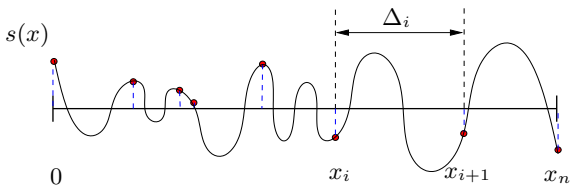}
    \end{psfrags}
}
\caption{Signal and sensor location}
\label{fig:sensorlocations}
\end{figure}

We consider in this paper the detection of a one dimensional
diffusion process that has a spatial state space structure as illustrated in
Fig.~\ref{fig:sensorlocations}.  Specifically,
we assume that the underlying signal $s(x)$ is
 the stationary solution\footnote{The stationary solution of
 (\ref{eq:diffusioneq}) requires a condition on  $A$, $B$,
 $\Pi_0$ given by $\Pi_0 = \frac{B^2}{2A}.$} of the diffusion equation
\begin{equation} \label{eq:diffusioneq}
\frac{ds(x)}{dx}= - A s(x)+ B u(x), ~~~x \ge 0,
\end{equation}
where  $A \ge 0$ and $B$ are known, and the initial condition is given by
$s(0) \sim \Nc(0, \Pi_0)$.
The input process $u(x)$ is  zero-mean white Gaussian, independent of
 both sensor noise  and $s(0)$.

We assume that each sensor in the field takes a noisy measurement.  At location $x_i$, the
sensor measurement $y_i$ comes from the following binary hypotheses
\begin{equation}  \label{eq:hypothesis}
\begin{array}{lcl}
H_0 &: & y_i = w_i, \\
H_1 &: & y_i = s_i+ w_i, ~~~~i=1,2,\cdots, n,\\
\end{array}
\end{equation}
where $s_i\defeq s(x_i)$, and $w_i$ are i.i.d. sensor measurement noises from $\Nc(0,\sigma^2)$ with a known variance $\sigma^2$.
We will not focus
here on minimizing the number of bits for the quantization  of $y_i$, nor will
we tackle the problem of how $y_i$ are sent to the fusion center for detection.
These are important design issues that must be treated separately.
Our focus in this paper is that,
if data are to be collected from the sensor field,
where they should be collected.  For energy efficiency,
we aim to find a scheduling scheme that require as few
samples as possible.

\vspace{-1em}
\subsection{Summary of Results}
\vspace{-0.5em}
We adopt the Neyman-Pearson formulation by fixing the detector size $\alpha$
and minimizing the miss  probability when the number $n$  of
samples is large.  Specifically, we are interested in two closely connected
design problems: the locations $\Xmsc=\{x_i\}$ where data should be
collected, and the rate of decay of miss probability
for detectors of size $\alpha$.  We will make the assumption that
samples are collected uniformly with equal spacing
$\Delta_i=\Delta$.  Such an idealized assumption can only be approximated in practice,
but it does provide an analytically tractable formulation that
leads to insights into energy efficient data collection.

The miss probability $P_M(\Delta,n;\alpha,\mbox{SNR})$ is
a function of sensor spacing $\Delta$, the sample size $n$ as well as
detector size $\alpha$ and SNR $\Gamma\defeq\frac{\Pi_0}{\sigma^2}$.
The energy efficient scheduling problem can then be
formulated as one of optimizing the error exponent $K_\alpha(\Delta,\mbox{SNR})$
under the large deviation principle where
\begin{equation}\label{eq:exp}
K_\alpha(\Delta,\Gamma)=\lim_{n\rightarrow \infty} \frac{1}{n}\log P_M(\Delta,n;\alpha,\Gamma).
\end{equation}
The connection of the above formulation
with energy efficient sensor scheduling is natural when the
sample size $n$ is directly related with the number of transmissions
(such as the case in Sensor Networks with Mobile Access (SENMA)).

We derive a closed-form expression for the error exponent
 $K_\alpha(\Delta,\Gamma)$
 of miss probability (which is independent of $\alpha$) by exploiting the
state-space structure of alternative hypotheses and making a connection
with Kalman filtering.  We show next that $K_\alpha(\Delta,\Gamma)$ has a distinct phase transition:
when SNR $\Gamma>1$, $K_\alpha(\Delta,\Gamma)$ is a monotonically
increasing function of $\Delta$, indicating that sensor spacing $\Delta$
 should be made as large as possible
(for fixed but large sample size).  When SNR $\Gamma<1$, on the other hand, $K_\alpha(\Delta,\Gamma)$ achieves the maximum at  some $\Delta^*$, which
means that there is an optimal spacing among collected samples. We also provide
 an implicit equation for  $\Delta^*$.  We also present simulation results
 to demonstrate the predicted behavior.

\vspace{-1em}

\vspace{-0.5em}
\subsection{Related Work}
The detection of Gauss-Markov process in Gaussian noise is a classical
problem. See \cite{Kailath&Poor:98IT} and references therein. Our work
relies on the  connection between likelihood ratio and the
innovation process through Kalman filtering by Schweppe \cite{Schweppe:65IT}.
While the connection between Kalman filter and error exponent
is a contribution of this paper, there is an extensive literature
on the large deviation approach to the detection Gauss-Markov process
\cite{Benitz&Bucklew:90IT, Bahr:90IT, Bahr&Bucklew:90SP, Barone&Gigli&Piccioni:95IT, Bryc&Smolenski:93SPL, Bercu&Gamboa&Rouault:97SPA, Luschgy:94SJS}.
  These results
do not provide explicit expressions (with an exception for noiseless AR processes)
from which optimal scheduling can be obtained.

The sensor scheduling problem can also be viewed as a sampling
problem. To this end, Bahr and Bucklew \cite{Bahr&Bucklew:90SP}
optimized the exponent numerically under a Bayesian formulation.
For a specific signal model (low pass signal in colored noise),
they showed that the optimal sampling depends on SNR, which we
also show in this paper in a different setting.

\vspace{-1.0em}
\section{System Model}
\label{sec:systemmodel}
\vspace{-0.5em}

We derive the discrete model for the sampled data of the signal
$s(x)$ similarly to \cite{Micheli&Jordan:02MTNS}. For equally
spaced sensors with spacing $\Delta$, the dynamics of sampled
signal are described by
\begin{eqnarray}
s_{i+1} &=& a s_i +  u_i, \label{eq:statespacemodel} \\
 a &=& e^{-A\Delta},~~
 u_i  = \int_0^{\Delta} e^{-Ax} B u(x_{i+1}-x) dx, \nn
\end{eqnarray}
The mean and variance of process  $u_i$ are given by
\beq
\Ebb u_i = 0, ~~~Q \defeq \Var (u_i) =\Pi_0(1-a^2).
\eeq
The signal samples $\{s_i\}$ form an autoregressive sequence under $H_1$.
Notice that $\Ebb s_i^2 = \Pi_0$ for all $i$ and the value of $a$ determines the amount of correlation between signal samples.  For i.i.d. signal samples we have $a=0$ and $a=1$ for the perfectly correlated signal.   Notice also that the noisy observation  $\{y_i\}$ are not autoregressive;
they follow the hidden Markov model.

\vspace{-1em}
\section{Error Exponent}
\vspace{-0.5em}

In this section, we investigate the performance of Neyman-Pearson detector with a level $\alpha \in (0, 1)$.   We present a closed-form expression of the {\em error exponent} of miss probability $K_\alpha(\Delta,\Gamma)$ defined
in (\ref{eq:exp}).
The theorem below comes from the fact that the limit of the normalized
log-likelihood ratio under $H_0$ is the best error exponent for general ergodic cases\cite{Vajda:book}.
By expressing the log-likelihood ratio through the innovation representation,
we make the closed-form calculation of error exponent tractable.

\begin{theorem}[Error exponent] \label{theo:errorexponentNP}
For the Neyman-Pearson detector  of the hypotheses (\ref{eq:hypothesis}, \ref{eq:statespacemodel}) with level $\alpha \in (0,1)$ (i.e. $P_F \le \alpha$)   and $0\le a \le 1$, the best error exponent of miss probability is independent of $\alpha$ and is given by
{\footnotesize
\begin{eqnarray}
K_\alpha(\Delta,\Gamma) &=&
-\frac{1}{2} \log\frac{\sigma^2}{R_{e}}
+ \frac{1}{2} \frac{\tilde{R}_{e}}{R_{e}}
 - \frac{1}{2}, \label{eq:errorexponent}
\end{eqnarray}
}
where $R_{e}$ and $\tilde{R}_{e}$ are the steady-state  variances of
the innovation process of $y_i$ calculated under $H_1$ and $H_0$, respectively.
Specifically, $R_{e}$ and $\tilde{R}_{e}$ are given by
{\scriptsize
\begin{eqnarray}
R_{e} &=& P + \sigma^2, \label{eq:Reinfexplicit}\\
\tilde{R}_{e} &=&\sigma^2\left(1+\frac{a^2P^2}{P^2+2\sigma^2P+(1-a^2)\sigma^4}\right),\label{eq:tReinfexplicit}\\
P &=& \frac{\sqrt{[\sigma^2(1-a^2)-Q]^2+4\sigma^2 Q}-\sigma^2(1-a^2)+Q}{2},\label{eq:Pinfexplicit}
\end{eqnarray}
}
In frequency domain,
\begin{equation}
K_\alpha(\Delta,\Gamma) = \frac{1}{2\pi}\int_0^{2\pi} D( \Nc(0,\sigma^2)||\Nc(0,S_y(\omega))) ~d\omega, \label{eq:errorexponentspectral}
\end{equation}
where $D(\cdot||\cdot)$ is the Kullback-Leibler distance, and the spectrum $S_y(\omega)$ of $\{y_i\}$ under $H_1$ is given by
\beq
S_y(\omega) = \sigma^2 + \frac{\Pi_0(1-a^2)}{1-2a\cos \omega + a^2}.
\eeq
\end{theorem}
Proof: see \cite{Sung&Tong&Poor:04ITsub}. $\blacksquare$

For notational convenience,  we use $K$ for the error exponent.

\vspace{-1em}
\subsection{Properties of Error Exponent }\label{subsec:errorexponent1D}

First,  it is easily seen from Theorem \ref{theo:errorexponentNP} that $K$ is a continuous function of the correlation coefficient $a$ ($0 \le a \le 1$) for a given $\Pi_0$ and $\sigma^2$ since $R_{e} \ge \sigma^2 > 0$.

\begin{theorem}
The error exponent  is positive for any SNR $\Gamma$ and $0 \le a < 1$.  Furthermore,
\begin{itemize}
\item[(i)] for i.i.d. observations ($a=0$), the error exponent  reduces to the Kullack-Leibler distance $D(p_0||p_1)$ where $p_0 \sim \Nc(0,\sigma^2)$ and $p_1 \sim \Nc(0, \Pi_0+\sigma^2)$;
\item[(ii)]
for Perfectly correlated signal ($a = 1$), the error exponent is zero
 for any SNR  $\Gamma$, and the miss probability is bounded by
 \beq \label{eq:PMCor}
(\frac{1}{\sqrt{2\pi}}-D)cn^{-1/2} \le P_M \le \frac{1}{\sqrt{2\pi}}cn^{-1/2}
\eeq
for sufficiently large $n$,
where $c$ and $D \in (0,\frac{1}{\sqrt{2\pi}})$ are positive constants.
\end{itemize}
\end{theorem}

The positivity of $K$ is immediate from Theorem~\ref{theo:errorexponentNP}.
The case when $a=0$ corresponds to the {\em Stein's lemma} for  the i.i.d. case. Unless $a=1$,
the miss detection probability always decays exponentially.

For the perfectly correlated case ($a=1$), the miss probability does not decay exponentially;
it decays with $\Theta(\frac{1}{\sqrt{n}})$ as shown in (\ref{eq:PMCor}). This
is explained by the form of the optimal detector. It can be shown that the  sufficient statistic is given by
\beq
T = |\sum_{i=1}^n y_i|^2.
\eeq
Under $H_0$, we have $\sum_{i=1}^n y_i \sim \Nc(0, ~n\sigma^2)$. Since $f(x)=x^2$ is symmetric about zero, the Neyman-Pearson detector with level $\alpha$ is simply given by
\beq \label{eq:detectorpercor}
\delta_n^{cor} = {\mathbf 1}_{\{|\sum_{i=1}^n y_i| ~\ge~ z_n\} },
\eeq
where $z_n = \sqrt{n}\sigma Q^{-1}(\frac{\alpha}{2})$. ($Q(\cdot)$ is the tail probability of standard normal distribution).  Since $P_{1,n} = \Nc(0, ~n^2 \Pi_0 + n\sigma^2)$, the probability of missing is given by
\beq
P_{M}^{} = 1 - 2 Q\left(\frac{\sqrt{n}\sigma Q^{-1}(\frac{\alpha}{2})}{\sqrt{n^2\Pi_0 + n\sigma^2}}\right).
\eeq
For large $n$, $P_M$ behaves as $1-2Q(\frac{c}{\sqrt{n}})$ where $c$ is a constant,
which leads to (\ref{eq:PMCor}).

Having obtained the behavior of error exponent at two extreme correlation cases,
we now show that the error exponent has distinct characteristics at different
SNR regimes.

\begin{theorem}[$K$ vs. correlation - high SNR ]  \label{theo:etavsahighsnr}
$K$ is monotone decreasing as the correlation strength increases (i.e. $a \uparrow 1$) if the SNR  $\Gamma > 1$.
\end{theorem}

\begin{figure}[htbp]
\centerline{
{
    \begin{psfrags}
    \epsfxsize=1.7in
    \epsfysize=1.2in
    \epsfbox{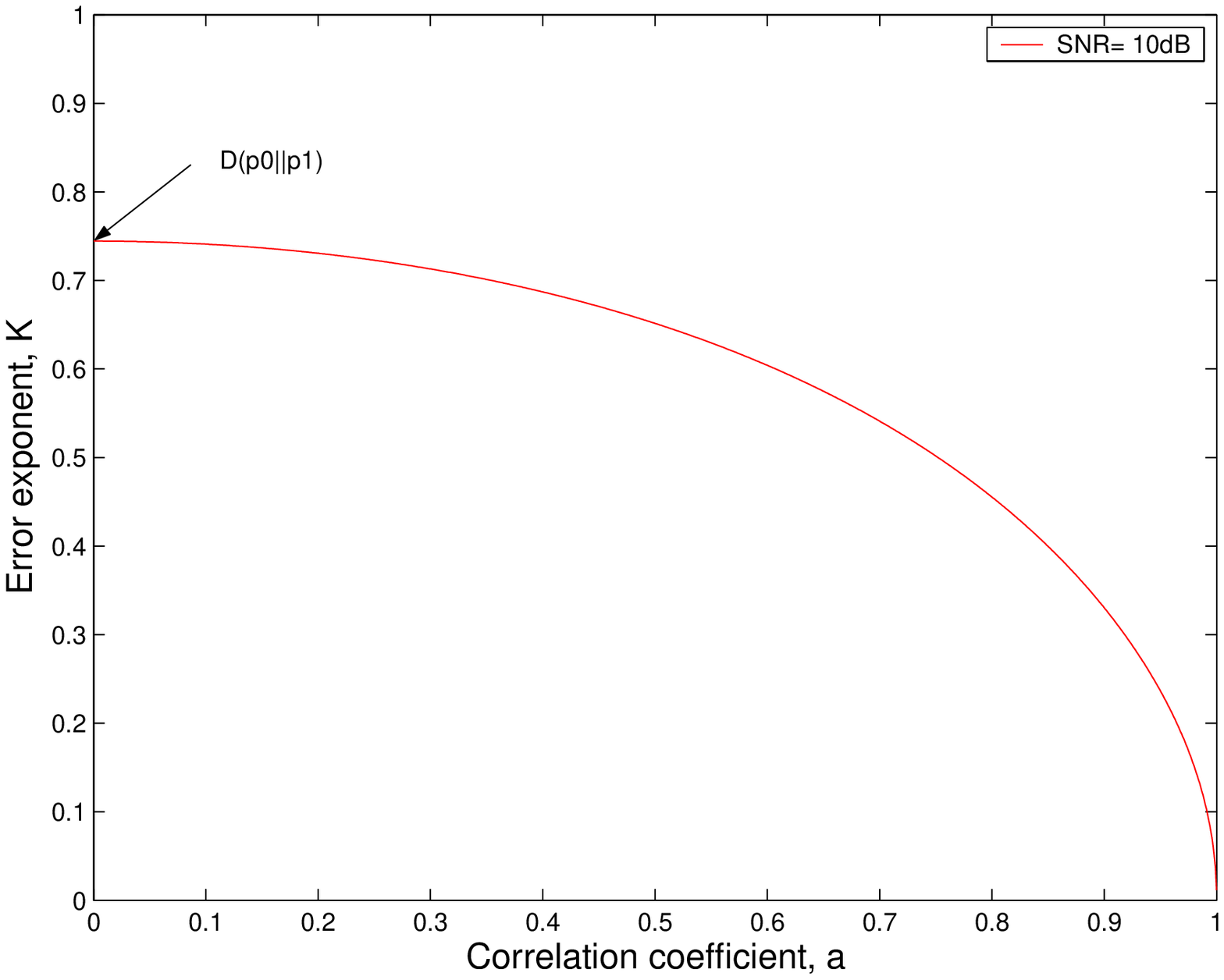}
    \epsfxsize=1.733in
    \epsfysize=1.2in
    \epsfbox{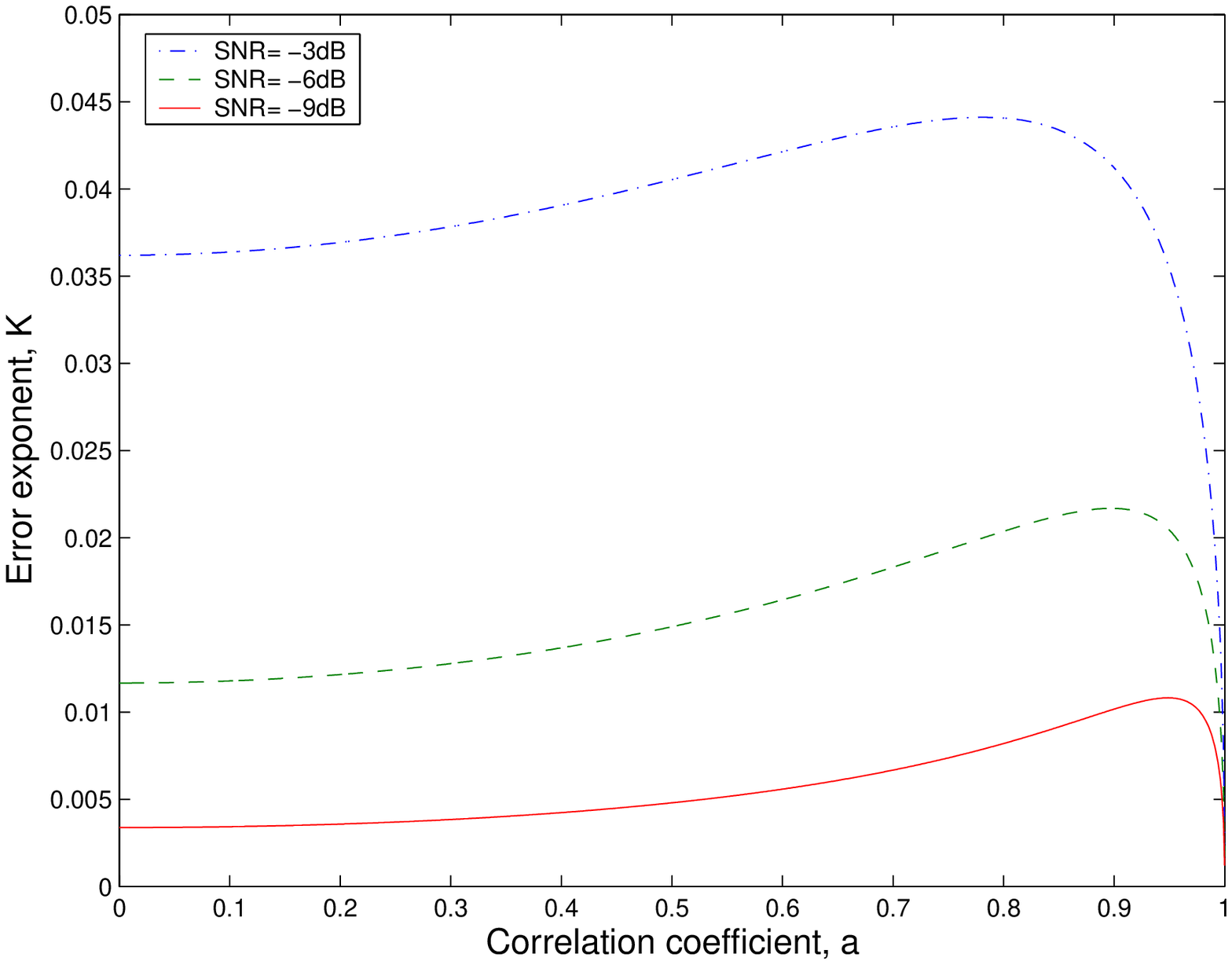}
    \end{psfrags}
}
}
\caption{$K$ vs. $a$ (left:SNR=10 dB, ~right:SNR=-3,-6,-9 dB)}
\label{fig:snr10dBetavsa}
\end{figure}

The above theorem shows that the i.i.d. observations give the best error performance for the same SNR with the maximum error exponent {\scriptsize $D(\Nc(0,1)||\Nc(0,1+\Gamma))$} when SNR is larger than unity.
Fig. \ref{fig:snr10dBetavsa} (left) shows the error exponent as a function of the correlation coefficient $a$. In the context sensor scheduling, if $n$ transmissions are required for the desired miss probability,
it is best to maximize the distances between the scheduled sensors.  The intuition is that, at
 high SNR, the signal component in the observation is strong, and the innovation process
contains more information about the hypotheses.  Thus there is benefit to make the
signal less correlated.

 In contrast, error exponent at low SNR is not monotonic, and there exists an optimal
 correlation.
Fig. \ref{fig:snr10dBetavsa} (right) shows the error exponent for the case that the SNR is less than unity.
As shown, the i.i.d. case is not the best for the error performance for the same SNR.  The error exponent initially increases as $a$ increases, and then decreases to zero as $a$ approaches one. For SNR of -6 dB the error exponent is less  than the case of -3 dB and it is seen that $a_m$ is shifted closer to one.
The following theorem characterizes the optimal correlation.

\begin{theorem}[$K$ vs. correlation - low SNR]  \label{theo:etavsalowsnr}
There exists a non-zero value $a_m$ of the correlation coefficient  that achieves the maximum  $K$ for $\mbox{SNR} < 1$, and $a_m$ is given
by solving the following equation.
\beq
[1+a^2+\Gamma(1-a^2)]^2-2(r_e+\frac{a^4}{r_e})=0,
\eeq
where $r_e = R_{e}/\sigma^2$.
Futhermore, $a_m$ converges to one as SNR $\Gamma$ goes to zero.
\end{theorem}

At low SNR, noise in the observation dominates.  Thus making signal more correlated
provides the benefit of noise averaging. The lower the SNR, the stronger the correlation
is desired as shown in Fig.~\ref{fig:snr10dBetavsa} (right).
Note however that excessive correlation in the signal doesn't provide
new information by the observation.  Notice also that the maximum value of error exponent is much larger than the value for i.i.d. case at lower SNR. Hence, the optimal correlation  gives much better performance with the same number of sensor observations for low SNR cases.

A significance of the above theorem is the determination of
optimal sensor spacing.  In particular, the optimal distance is given by
\beq
\Delta^* = -\log (a_m) / A,
\eeq
for the same underlying physical phenomenon described by (\ref{eq:diffusioneq}).

\begin{figure}[htbp]
\centerline{
{
    \begin{psfrags}
    \epsfxsize=1.7in
    \epsfysize=1.2in
    \epsfbox{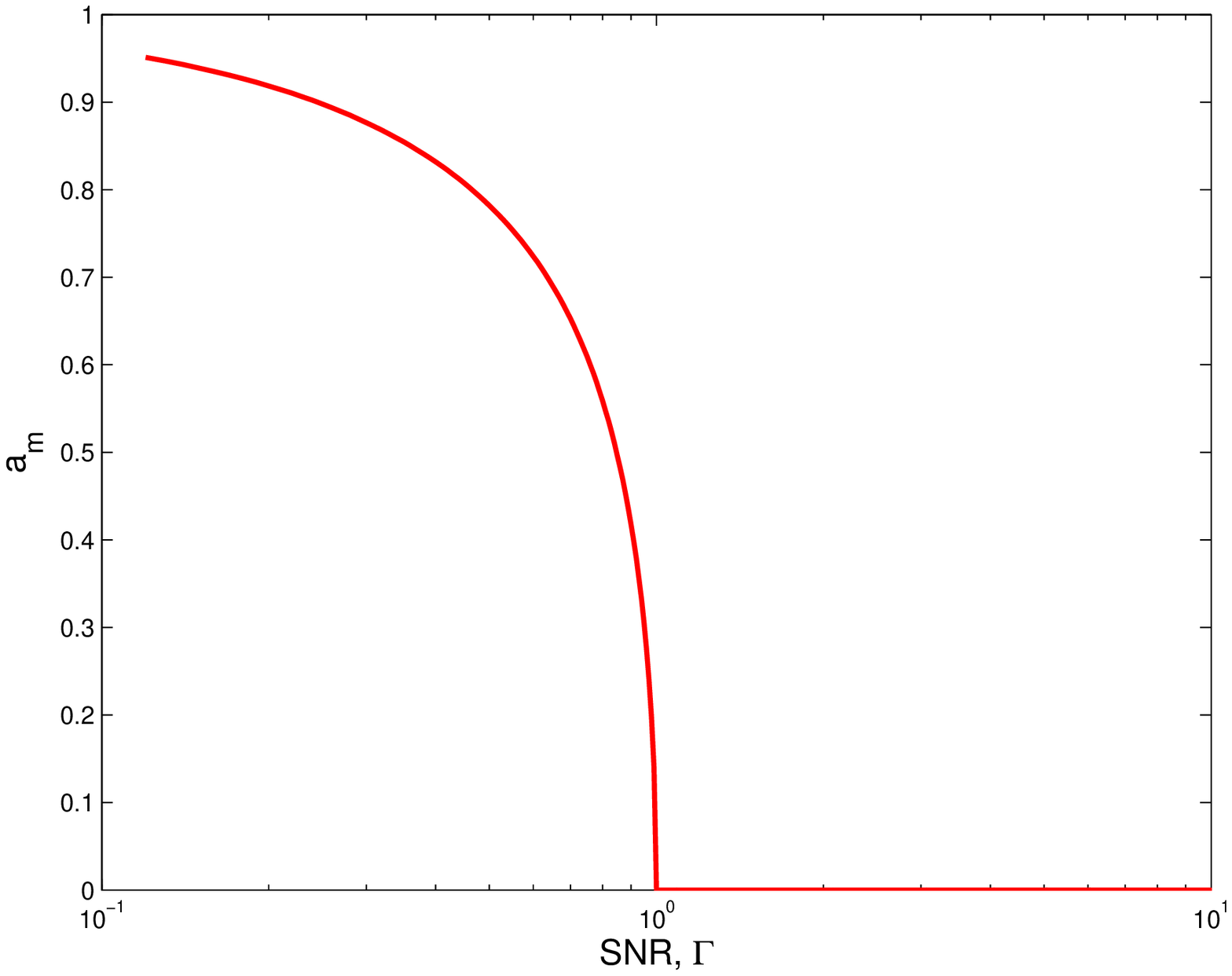}
    \epsfxsize=1.7in
    \epsfysize=1.2in
    \epsfbox{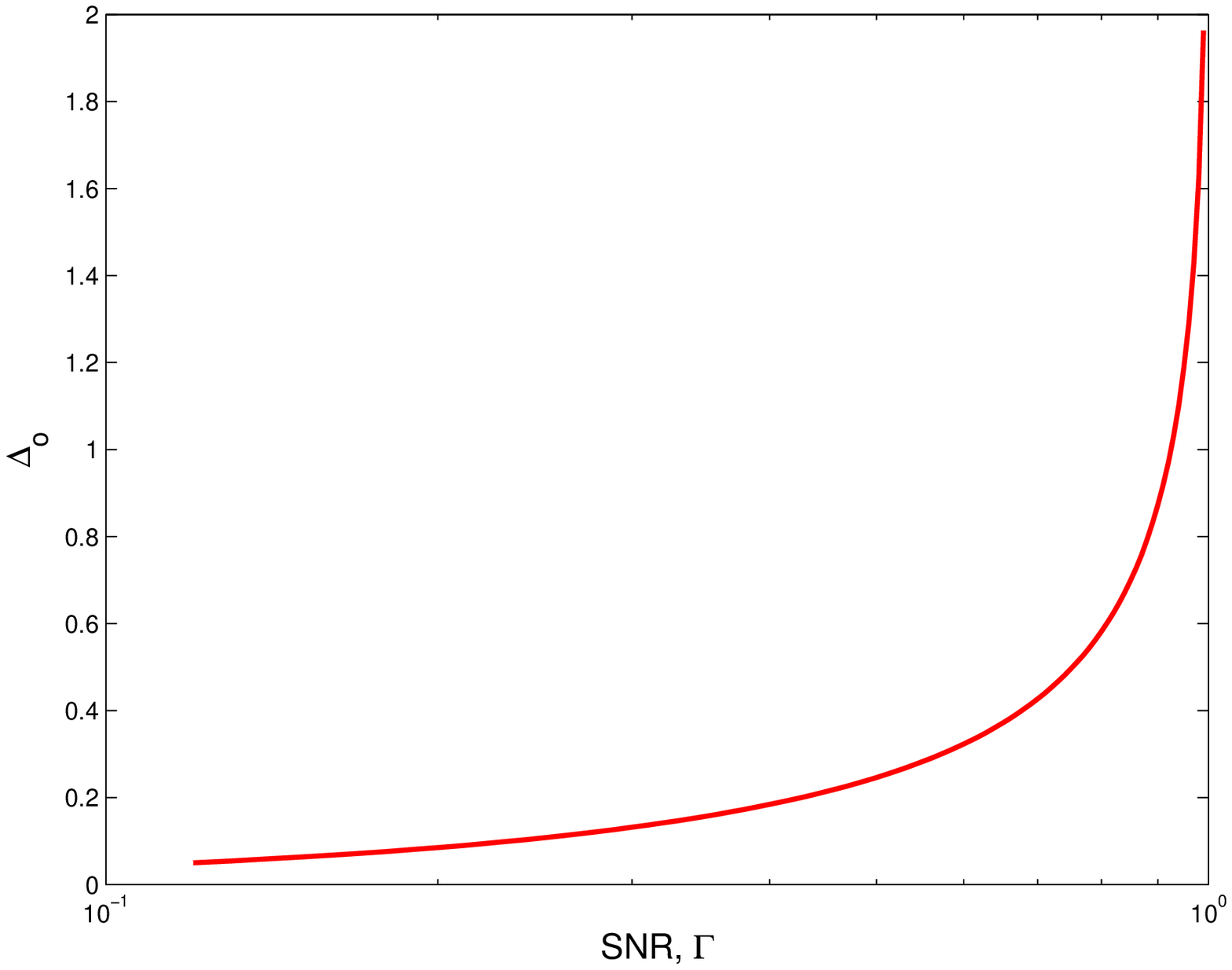}
    \end{psfrags}
}
}
\caption{Left: $a_m$ vs $\Gamma$, ~~right:$\Delta^*$ vs $\Gamma ( < 1)$}
\label{fig:amaxvssnr}
\end{figure}

Fig. \ref{fig:amaxvssnr} shows the value of  $a_m$ that maximizes the error exponent as a function of SNR.  As shown in the figure  the SNR of {\em unity} gives sharp transition between two different behaviors of error  exponent w.r.t. correlation strength.

 Finally, we investigate the behavior of the error exponent w.r.t. SNR.
\begin{theorem}[$K$  vs SNR] \label{theo:etavsSNR}
The error exponent  $K$ is monotone increasing as SNR increases for
a given correlation coefficient $0 \le a < 1$.  Moreover, at high SNR $K$ increases linearly with
 $\log \mbox{SNR}$.
\end{theorem}
\begin{figure}[htbp]
\centerline{
{
    \begin{psfrags}
    \epsfxsize=2.0in
    \epsfysize=1.4in
    \epsfbox{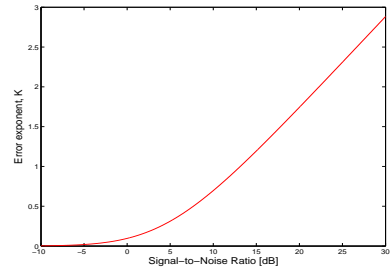}
    \end{psfrags}
}
}
\caption{$K$ versus SNR ($a=e^{-1}$)}
\label{fig:etavsSNR1}
\end{figure}
The $\log$ SNR increase of $K$ w.r.t. SNR is similar to the case of diversity combining of Rayleigh-faded multipaths in additive noise since in both cases the signal component is random.  Comparing with the detection of a deterministic signal in noise where the error exponent is proportional to SNR,  the increase of error exponent w.r.t. SNR is much slower for the case of stochastic signal in noise.

\subsection{Sensor Placement}
So far we have assumed that sensors have been placed and the problem is to schedule
the transmission of selected sensors.  A related problem is sensor placement.
Should $n$ sensors be placed to cover as large an area as possible, or should
they be clustered in a subregion of the signal field?
 Theorem~\ref{theo:etavsahighsnr}
suggests that, at high SNR,  these sensors should be placed with maximum separation subject to the size of the field.  At low SNR, however, sensors should be placed with the optimal separation.  This implies
that, if the sensor field is large, it is better to cluster the sensor in some region, leaving
other areas without sensors.

\vspace{-1em}
\section{Simulation Results}
\vspace{-0.7em}
\label{sec:numerical}

We considered the Neyman-Pearson detector of first-order autoregressive signal described by (\ref{eq:statespacemodel}). We considered SNR of 10 dB and - 3 dB, and several $a$ for each SNR. The probability of false alarm was set 0.1\% for all cases.
\vspace{1.8em}
\begin{figure}[htbp]
\centerline{
    \begin{psfrags}
    \epsfxsize=2.0in
    \epsfysize=1.4in
    \epsfbox{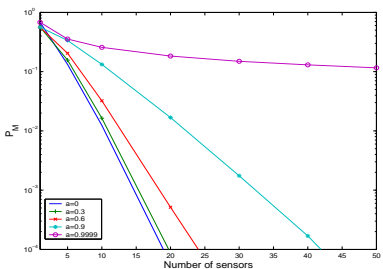}
    \end{psfrags}
}
\caption{ $P_M$ vs.  \# of sensors (SNR=10dB)}
\label{fig:PMvsnosensors10dB}
\end{figure}
Fig. \ref{fig:PMvsnosensors10dB} shows the miss probability w.r.t. number of sensors for 10 dB SNR.  It is shown that the i.i.d. case ($a=0$)  has the largest slope for error performance, and the slope of error decay is monotonically decreasing as $a$ increases to one. Notice that the error performance for the same number of sensors is significantly different for different correlation strength even for the same SNR, and the performance for weak correlation is not much different from i.i.d. case predicted by Fig. \ref{fig:snr10dBetavsa} (left). (We can see that the slope decreases suddenly near $a=1$.) It is shown that the miss probability for highly correlated case ($a=0.9999$) is not exponenitally decay.
\begin{figure}[htbp]
\centerline{
    \begin{psfrags}
    \epsfxsize=2.0in
    \epsfysize=1.4in
    \epsfbox{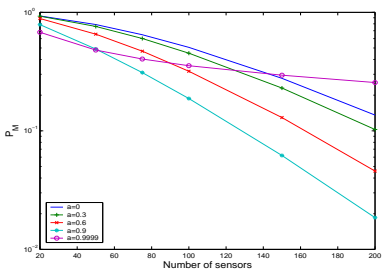}
    \end{psfrags}
}
\caption{ $P_M$ vs.  \# of sensors (SNR=-3dB)}
\label{fig:PMvsnosensorsm3dB}
\end{figure}

The error performance for SNR of -3 dB is shown in Fig. \ref{fig:PMvsnosensorsm3dB}. It is seen that the slope increases as $a$ increases from zero, and reaches maximum  with sudden decrease after the maximum. Notice that the error curve is still not a straight line for low SNR case due to the $o(n)$ term in the exponent.  Since the error exponent increases only with $\log$ SNR, the required number of sensors for -3 dB SNR  is much larger than for 10 dB SNR for the same miss probability. It is clearly seen that $P_M$ is still larger than $10^{-2}$  for 200 sensors whereas it is $10^{-4}$ with 20 sensors for 10 dB SNR case.

\section{Conclusions}
\label{sec:conclusion}

We have considered the detection of correlated signal
using  noisy sensors.  We have derived the best error exponent for
 the Neyman-Pearson detector satisfying a fixed size constraint
using the innovations and the spectral domain approaches. We have
also investigated
the properties of the error exponent.
The error exponent is a function of SNR and correlation strength.
The behavior of error exponent w.r.t. correlation strength
is sharply divided to two regions depending on SNR. For SNR larger than unity
the error exponent is monotone decreasing as correlation becomes strong. On the
other hand it has a non-i.i.d. correlation strength that gives the maximum slope
for SNR smaller than one. Using the property of error exponent,
the optimal strategy for sensor scheduling  has been derived.

The results presented in this paper have a generalization to the
vector state-space model.  Such a generalization is useful to consider
 more general scheduling schemes.
In practice, the locations of sensors are random, and sensors
may cease to function.  Thus random sampling may need to be considered.


{\scriptsize
\bibliographystyle{plain}

}

\end{document}